\documentclass[11pt]{article}

\usepackage{latexsym}
\usepackage{amssymb}
\usepackage{epsf}
\usepackage{amsmath}
\usepackage[hypertex]{hyperref}
\usepackage{graphicx}

\begin{document}
\begin{flushright}
\vspace*{-1cm}
LA-UR-07-7938\\
SLAC-PUB-13059
\end{flushright}

\begin{center}{\Large \bf
Supersymmetric Model Building\\
(and Sweet Spot Supersymmetry)\footnote{ Based on lectures given by
 R.~Kitano at the Summer Institute 2007, Fuji-Yoshida, Japan.  Most of
 the content is based on works done at the Stanford Linear Accelerator
 Center under support of the U.S. Department of Energy (contract number
 DE-AC02-76SF00515).} }
\end{center}

\begin{center}
Masahiro Ibe$^{a,b}$
and Ryuichiro Kitano$^c$
\vspace{6pt}\\

$^a${\it Stanford Linear Accelerator Center, Stanford University,
                Stanford, CA 94309} \\
$^b${\it Physics Department, Stanford University, Stanford, CA 94305}\\
$^c${\it 
Theoretical Division T-8, Los Alamos National Laboratory, Los Alamos, NM 87545
}\\
\end{center}
\begin{abstract}
It has been more than twenty years since theorists started discussing
  supersymmetric model building/phenomenology. We review mechanisms of
  supersymmetry breaking/mediation and problems in each scenario. We
  propose a simple model to address those problems and discuss its
  phenomenology.
\end{abstract}

\section{Introduction}

\setcounter{footnote}{0}

There are three questions we should ask when we try to construct a model
with weak scale supersymmetry; (1) How supersymmetry is broken? (2) How
gauge/matter fields feel the supersymmetry breaking? (3) How the Higgs
fields feel the supersymmetry breaking? There is a simple mathematical
formulation of those questions. It is all about the interaction terms in
the Lagrangian among the Goldstino multiplet $S$ and the fields in the
minimal supersymmetric standard model (MSSM).

We reconsider problems in supersymmetric models by using the
effective-field-theory approach. The low energy theory is described by
the field $S$ and the matter/gauge/Higgs fields. By doing so, we can
discuss different scenarios as different choices of functional form of
$S$ which define an effective theory.
In this formulation, we find that there is a sweet spot in between the
gauge and gravity mediation ($m_{3/2} \sim O(1)$~GeV) where the theory
is perfectly consistent with various requirements.

This note is based
on~\cite{Kitano:2006wm,Kitano:2006wz,Ibe:2006rc,Ibe:2007km}.

\section{Formulation}

\setcounter{footnote}{0}

\subsection{Question (1): Supersymmetry Breaking Sector}
\label{sec:s-sector}

The construction of the low-energy effective Lagrangian for
supersymmetry breaking sector is completely analogous to the Higgs
sector in the standard model. In the standard model, the Higgs potential
is given by
\begin{eqnarray}
 V = {\lambda_H \over 4} \left( |H|^2 - v^2 \right)^2\ .
\end{eqnarray}
Here the Higgs field $H$ is a linear representation of the gauge group,
and can be decomposed to the vacuum expectation value (VEV), three
Goldstone bosons and a physical field. This theory makes sense as a
low-energy effective theory as long as $\lambda_H \lesssim 4 \pi$. Not
only as the simplest model of the electroweak symmetry breaking, this
model serves as the effective theory of a wide class of models which
arises after integrating out heavy fields. The two parameters
$\lambda_H$ and $v$ respectively set the strength of the
self-interaction of the Higgs fields and the electroweak scale. The
physical mode and gauge bosons get masses $\sqrt{\lambda_H } v$ and $g
v$, respectively.

\begin{figure}[t]
\begin{center}
  \includegraphics[width=8.5cm]{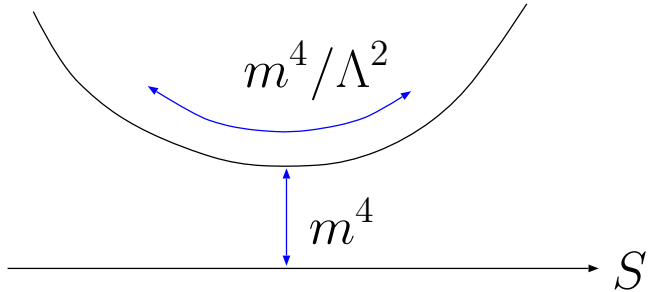}
\end{center}
\caption{Any supersymmetry breaking model locally looks like this. We
 can parametrize this potential by two parameters $m^2$ and
 $\Lambda^2$. If it is a model of supersymmetry breaking by an $F$-term
 and if $m^4/\Lambda^4 \lesssim 4 \pi$, the effective Lagrangian to
 describe physics around the minimum is given by
 Eq.~(\ref{eq:susy-breaking}).}
\label{fig:susyB}
\end{figure}

A wide class of supersymmetry breaking models also has a low-energy
effective description in terms of the Goldstino multiplet $S$, that
contains the $F$-component VEV, the Goldstino fermion and a physical
scalar field. The Lagrangian is defined by a K\"ahler- and a
superpotential as follows:
\begin{eqnarray}
 K = S^\dagger S - \frac{(S^\dagger
  S)^2}{\Lambda^2} \ ,\ \  W = m^2 S \ .
\label{eq:susy-breaking}
\end{eqnarray}
We eliminated the possible cubic terms $S^\dagger S^2 + {\rm h.c.}$ by
an appropriate field redefinition so that $S=0$ is the minimum of the
potential. This theory makes sense if $m^2 \lesssim \sqrt{4 \pi}
\Lambda^2$. The two parameters $1/\Lambda^2$ and $m^2$ respectively set
the strength of the self-interaction of the $S$ field and the
supersymmetry breaking scale $F_S$. The physical mode and the gravitino
obtain masses: $m_S = 2 m^2 / \Lambda $ and $m_{3/2} = m^2 / \sqrt 3 M_{\rm
Pl}$.

Since we have the standard model of supersymmetry breaking in
Eq.~(\ref{eq:susy-breaking}), we do not need to consider detail of
mechanisms or dynamics of the supersymmetry breaking for now. Each model
of supersymmetry breaking simply maps to a parameter point or region in
the two-dimensional parameter space ($m_{3/2}, \Lambda$). We can leave
those as parameters in the low-energy theory.

\subsection{Minimal Supersymmetric Standard Model}
\label{sec:mssm}

Before going to the discussion of the communication between the MSSM
sector and the supersymmetric sector, we review here important features
of the MSSM.

The superpotential of the MSSM is
\begin{eqnarray}
 W_{\rm MSSM} = Q H_u U + Q H_d D + L H_d E + \mu H_u H_d\ ,
\end{eqnarray}
where we suppressed the Yukawa coupling constants and flavor indices.
The last term, the $\mu$-term, is needed to give a mass to the Higgsino,
but it should not be too large. For supersymmetry to be a solution to
the hierarchy problem, i.e., $\langle H_{u,d} \rangle \ll M_{\rm Pl}$, the
$\mu$-term is necessary to be of the order of the electroweak scale (or
scale of the soft supersymmetry breaking terms).

This is called the $\mu$-problem. The fact that $\mu$ is much smaller
than the Planck scale suggests that the combination of $H_u H_d$
carries some approximately conserving charge.

There are many gauge invariant operators we can write down in addition
to the above superpotential such as
\begin{eqnarray}
 W_{R \!\!\! /} =
UDD + LLE + QLD \ ,
\label{eq:r-breaking}
\end{eqnarray}
and
\begin{eqnarray}
 W_{\rm dim. 5} = QQQL + UDUE\ .
\end{eqnarray}
These are unwanted operators as they cause too rapid proton decays.

The $\mu$-problem and the proton decay problem above are actually
related, and there is a simple solution to both problems. The
Peccei-Quinn (PQ) symmetry with the following charge assignment avoids
too large $\mu$-term and the proton decay operators.
\begin{eqnarray}
 PQ(Q) = PQ(U) = PQ(D) = PQ(L) = PQ(E) = - \frac{1}{2}\ ,
\end{eqnarray}
\begin{eqnarray}
 PQ(H_u) = PQ(H_d) = 1\ .
\end{eqnarray}
This symmetry is broken explicitly by the $\mu$-term, $PQ(\mu)=-2$.
Since it is a small breaking of the PQ symmetry, the coefficients of the
dimension five operators are sufficiently suppressed.
The unbroken $Z_4$ symmetry, which includes the $R$-parity as a
subgroup, still forbids the superpotential terms in
Eq.~(\ref{eq:r-breaking}) and ensures the stability of the lightest
supersymmetric particle (LSP), leaving us to have a candidate for dark
matter of the universe.

In fact, there is another symmetry which can play the same role as the
PQ symmetry, called $R$-symmetry. The charge assignment is
\begin{eqnarray}
 R(Q) = R(U) = R(D) = R(L) = R(E) = 1\ ,
\end{eqnarray}
\begin{eqnarray}
 R(H_u) = R(H_d) = 0\ .
\end{eqnarray}
Again, $R(\mu) = 2$ explicitly breaks the $R$-symmetry down to the
$R$-parity. 

In summary, there are approximate symmetries, U(1)$_{PQ}$ and U(1)$_R$,
in the Lagrangian of the MSSM.  If one of them is a good (approximate)
symmetry of the whole system, it provides us with a solution to the
$\mu$ and the proton decay problems.

\subsection{Question (2): Interactions between \boldmath{$S$} and gauge/matter fields}
Now we discuss interaction terms between the $S$-sector and the MSSM
sector. These interactions determine the pattern of supersymmetry
breaking parameters which are relevant for low energy physics.

There are actually only three possibilities for gauginos to obtain
Majorana masses. The Majorana mass of the gauginos is given by
\begin{eqnarray}
 m_{1/2} = {[f]_F \over [f]_A}\ ,
\label{eq:majorana}
\end{eqnarray}
where $f$ is the gauge kinetic function, ${\cal L} \ni f W^\alpha
W_\alpha$. The gauge kinetic function $f$ is a one-loop exact quantity:
\begin{eqnarray}
 f = {1 \over g^2(\Lambda_0)} 
- \sum_i {2 b_H^i \over (4 \pi)^2 } \log {M_H^i \over
  \Lambda_0}
- {2 b_L \over (4 \pi)^2 } \log {\mu_R \over \Lambda_0}\ ,
\end{eqnarray}
where $b_H^i$ and $M_H^i$ are the contributions to the beta-function and
masses of heavy fields in the theory, respectively. The first term is
the gauge coupling constant at tree level (defined by the running
coupling constant at a cut-off scale $\Lambda_0$). The last term is the
contribution from one-loop correction by light fields.
In order for the function $f$ to be $S$ dependent so that r.h.s. of
Eq.~(\ref{eq:majorana}) is non-vanishing~\cite{Giudice:1997ni}, we have
three choices, i.e., which term(s) is $S$ dependent?
Each of possibilities has been named as the
gravity~\cite{Chamseddine:1982jx},
gauge~\cite{Dine:1981za,Dine:1993yw,Dine:1994vc} and anomaly
mediation~\cite{Randall:1998uk}, respectively.

The tree-level gauge coupling $g$ can be $S$ dependent if
\begin{eqnarray}
 f(S)_{\rm tree} = {1 \over g^2} + {S \over M_{\rm Pl}} + \cdots\ .
\end{eqnarray}
We call this assumption as `gravity mediation.' An important feature
here is that the $S$ field must be singlet under any symmetry even
including approximate ones at low energy.
Symmetries are either absent or badly broken in order to obtain $O(1)$
valued gauge coupling constants for the standard model gauge
interactions. Therefore, there is generically a cosmological moduli
problem in this scenario~\cite{Coughlan:1983ci}. Also, the absence of
(approximate) symmetry makes the supersymmetric CP problem sharper as we
see later. The gaugino masses are
\begin{eqnarray}
 M_{1/2} \sim {F_S \over M_{\rm Pl}} \sim m_{3/2}.
\end{eqnarray}
The gravitino mass $m_{3/2}$ is $O(100~{\rm GeV})$ in this scenario.

The second possibility, making the second term $S$-dependent, is called
`gauge mediation.' The unique possibility of making this term
$S$-dependent is
\begin{eqnarray}
 M_H^i \to M_H^i (S)\ .
\end{eqnarray}
The heavy fields whose masses depend on $S$ are called messenger
fields. The simplest possibility is to take
\begin{eqnarray}
 M_H = k S\ ,
\label{eq:mheavy}
\end{eqnarray}
where $k$ is a dimensionless constant. In this case, in contrast to
gravity mediation, $S$ can carry a charge since $S \to S e^{i \alpha}$
would only shift the $\theta$-term leaving the gauge coupling constant
unchanged. We can, therefore, assign a (anomalous) U(1) charge to the
$S$ field. (The $m^2 S$ term in Eq.~(\ref{eq:susy-breaking}) breaks this
symmetry explicitly. This explains the smallness of the scale of
supersymmetry breaking $m^2 \ll \Lambda^2$ which is consistent with the
assumption made before ($m^2 \lesssim \sqrt{4 \pi} \Lambda^2$). For
small values of $m^2$, we can regard the U(1) symmetry as an approximate
symmetry of the Lagrangian.) The gaugino
masses are~\cite{Dine:1993yw,Dine:1994vc}
\begin{eqnarray}
 M_{1/2} = {g^2 b_H \over (4 \pi)^2} {F_S \over \langle S \rangle}
= {g^2 b_H \over (4 \pi)^2} m_{3/2} 
\left( 
{\langle S \rangle \over \sqrt 3 M_{\rm Pl}} 
\right)^{-1}\ .
\label{eq:gaugino-gm}
\end{eqnarray}
The gravitino mass is $m_{3/2} \ll O(100~{\rm GeV})$ in this
scenario. Note that $S$ must have a VEV to be consistent if we take the
assumption in Eq.~(\ref{eq:mheavy}).

The third possibility is called `anomaly mediation.' In the supergravity
Lagrangian with a requirement that the cosmological constant is
vanishing, dimensionful parameters in the Lagrangian are accompanied
with the compensator field $\phi = 1 + m_{3/2} \theta^2$, where
$m_{3/2}\equiv F_S / \sqrt 3 M_{\rm Pl}$. Therefore, the third term
effectively becomes $S$-dependent:
\begin{eqnarray}
 - {2 b_L \over (4 \pi)^2 } \log {\mu_R \over \Lambda_0 \phi}\ .
\end{eqnarray}
The gaugino masses are
\begin{eqnarray}
 M_{1/2} = {g^2 b_L \over (4 \pi)^2} m_{3/2} \ .
\end{eqnarray}
Therefore, $m_{3/2} \sim 10 - 100~{\rm TeV}$ in this scenario.

What we can do is to choose at least one of the above three for the
functional form of the gauge kinetic function $f(S)$. For matter fields,
we need to choose wave function factors $Z_\Phi (S,S^\dagger)$ where
$K \ni Z_\Phi (S,S^\dagger) \Phi^\dagger \Phi$ and $\Phi = Q,U,D,L,E$.

\subsection{Question (3): Interactions between \boldmath{$S$} and the
  Higgs fields}

The interaction terms we can write down are
\begin{eqnarray}
K_{\rm Higgs} = 
Z_{H_u} (S,S^\dagger) H_u^\dagger H_u
+ Z_{H_d} (S,S^\dagger) H_d^\dagger H_d
+ Z_{H_u H_d} (S,S^\dagger) H_u H_d + {\rm h.c.}
\end{eqnarray}
We expand $Z$-functions as 
\begin{eqnarray}
 Z_{H_u} = 1 + (a_1 S + a_1^* S^\dagger) + a_2 S^\dagger S + \cdots,
\end{eqnarray}
\begin{eqnarray}
 Z_{H_d} = 1 + (b_1 S + b_1^* S^\dagger) + b_2 S^\dagger S + \cdots,
\end{eqnarray}
\begin{eqnarray}
 Z_{H_u H_d} = c_0 + (c_1 S + c_1^* S^\dagger) + c_2 S^\dagger S + \cdots,
\end{eqnarray}
where we have rescaled the fields $H_u$ and $H_d$ such that the kinetic
terms are canonically normalized. By an appropriate shift we also take
$\langle S \rangle = 0$. The coefficients $a_1$, $b_1$, $c_0$, $c_1$ and
$c_2$ are complex parameter whereas $a_2$ and $b_2$ are real.
Then soft terms are obtained to be
\begin{eqnarray}
 \mu = c_1^* F_S^\dagger + c_0 m_{3/2}\ ,
\end{eqnarray}
\begin{eqnarray}
 B \mu = c_2 |F_S|^2 - (a_1+b_1) F_S \mu + m_{3/2} \mu\ ,
\label{eq:bmu-general}
\end{eqnarray}
\begin{eqnarray}
 m_{H_u}^2 = -(a_2 - |a_1|^2) |F_S|^2 + O(m_{3/2})\ ,
\end{eqnarray}
\begin{eqnarray}
 m_{H_d}^2 = -(b_2 - |b_1|^2) |F_S|^2 + O(m_{3/2})\ .
\end{eqnarray}
$F_S$ and $m_{3/2}$ are related by $F_S = \sqrt{3} m_{3/2} M_{\rm Pl}$.

For natural electroweak symmetry breaking, we need
\begin{eqnarray}
 \mu^2 \sim m_{H_u}^2 \sim m_{H_d}^2 \sim m_W^2,
\label{eq:relation}
\end{eqnarray}
and
\begin{eqnarray}
 B \mu \lesssim m_W^2 \ ,
\label{eq:bmu}
\end{eqnarray}
where $m_W$ is the electroweak scale $(O(100~{\rm GeV}))$.
Also, to suppress CP violation at low energy we need
\begin{eqnarray}
 {\rm arg} (M_{1/2} \mu (B \mu)^*) \ll 1 \ .
\end{eqnarray}

In gravity mediation, all the coefficients are expected to be non-zero
and $O(1)$ in the unit of $M_{\rm Pl} = 1$ since we cannot assign any
charge to $S$.
The relation in Eq.~(\ref{eq:relation}) can be naturally explained in
this case. This is the Giudice-Masiero mechanism in gravity
mediation~\cite{Giudice:1988yz}.
However, at the same time, there is no reason to expect the CP phase
above to be small.

The simplest possibility of realizing a small phase is to take
\begin{eqnarray}
 B \mu \ll m_W^2
\end{eqnarray}
at an energy scale. 
This is possible if we can arrange 
\begin{eqnarray}
 a_1 = b_1 = c_2 = 0\ ,\ \ m_{3/2} \ll c_1 F_S\ .
\end{eqnarray}
The relation in Eq.~(\ref{eq:relation}) implies
\begin{eqnarray}
 c_1^2 \sim a_2 \sim b_2\ ,\ \  c_1 F_S \sim m_W\ .
\end{eqnarray}

The above general discussion is suggesting that $S$ field carries some
approximately conserving charge. If we assign the PQ-charge
\begin{eqnarray}
 PQ(S) = 2\ ,
\end{eqnarray}
we can naturally explain $a_1 = b_1 = c_2 = 0 (= c_0)$. Gauge mediation
is consistent with this hypothesis for generating gaugino masses because
$S$ can carry an anomalous U(1) charge and $m_{3/2}$ is small. Note that
anomaly mediation predicts $B\mu / \mu = m_{3/2} \sim 10-100$~TeV (see
the last term in Eq~.(\ref{eq:bmu-general})) which is inconsistent with
Eqs.~(\ref{eq:relation}) and (\ref{eq:bmu}). If $S$ carries the
PQ-charge, the forms of the $Z$-functions are
\begin{eqnarray}
 Z_{H_u} = 1 + {S^\dagger S \over \Lambda_H^2}\ ,\ \ 
 Z_{H_d} = 1 + {S^\dagger S \over \Lambda_H^2}\ ,\ \ 
 Z_{H_u H_d} = {S^\dagger \over \Lambda_H}\ ,
\end{eqnarray}
with $O(1)$ coefficients. The size of $\Lambda_H$ should be such that
\begin{eqnarray}
 \mu \sim {F_S \over \Lambda_H} \sim O(100~{\rm GeV})\ .
\label{eq:mu-gm}
\end{eqnarray}
In gauge mediation, this means $\Lambda_H \ll M_{\rm Pl}$. Therefore,
the general picture we obtain is that {\it supersymmetry breaking sector
and the Higgs fields are directly coupled above some energy scale
$\Lambda_H$ and the combined sector should respect the approximate
PQ-symmetry}.

A framework with gauge mediation $+$ approximate PQ symmetry $+$ the
Giudice-Masiero mechanism seems to be a good starting point according to
the discussion of the $\mu$-term, proton decays and the CP
problem. However, there is still an unexplained coincidence. For the
gaugino masses in Eq.~(\ref{eq:gaugino-gm}) and the $\mu$-term in
Eq.~(\ref{eq:mu-gm}) to be both $O(100~{\rm GeV})$, we need to explain a
relation:
\begin{eqnarray}
 \langle S \rangle \sim {\Lambda_H \over 100}\ .
\label{eq:100}
\end{eqnarray}
If there is a reasonable explanation for this coincidence in a
stabilization mechanism of $S$, this hypothesis is going to be a perfect
framework.
We summarize three mediation mechanisms in Table~\ref{tab:scenarios}. 

\begin{table}[t]
 \begin{center}
  \includegraphics[width=15cm]{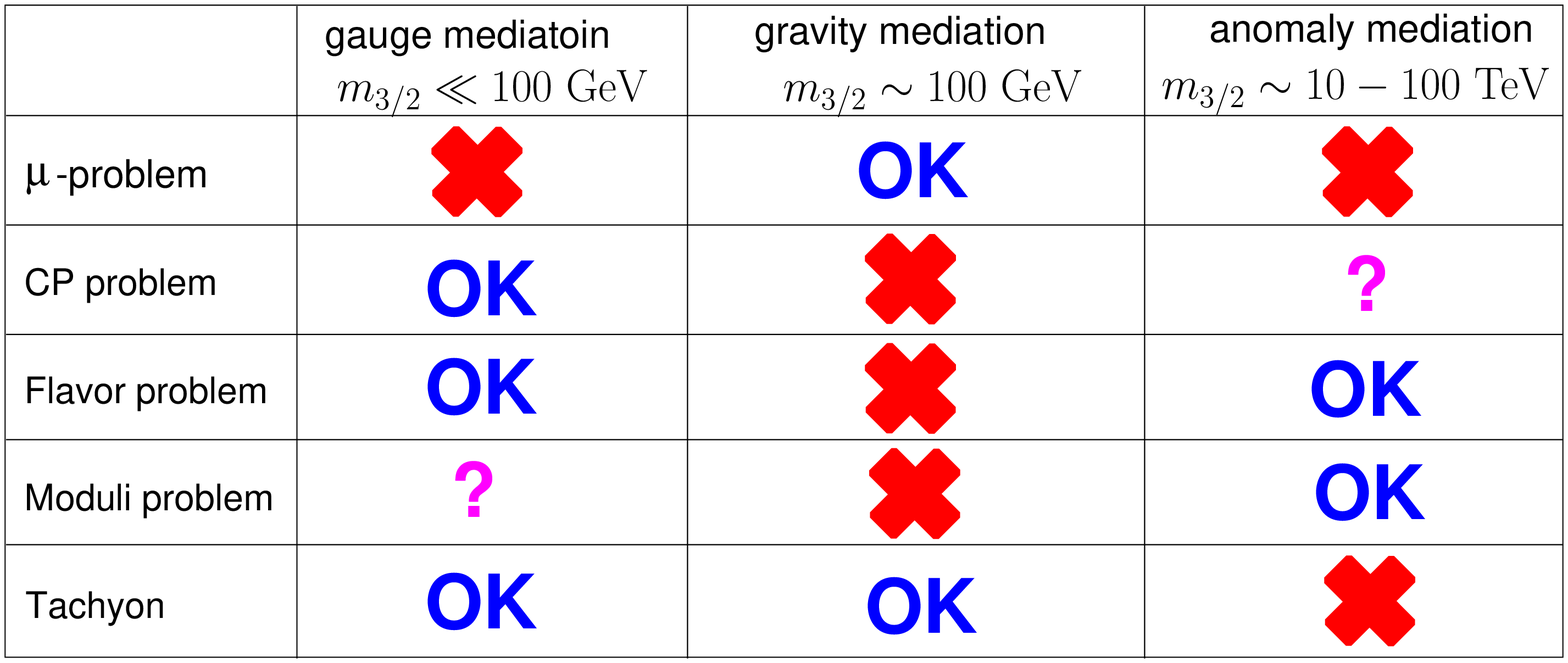}
\end{center}
\caption{Problems in each scenario of mediation mechanisms. We have to
 choose at least one of the mechanisms to obtain Majorana masses for
 gauginos.} \label{tab:scenarios}
\end{table}

\section{Sweet Spot Supersymmetry}
\label{sec:sss}

There have been many attempts to circumvent problems summarized in
Table~{\ref{tab:scenarios}}. For example, in
Ref.~\cite{Dine:1993yw,deGouvea:1997cx} it has been proposed to extend a
model of gauge mediation to the NMSSM by introducing a new singlet
field. However, for the successful electroweak symmetry breaking,
further extension of the model were necessary such as introduction of
vector-like matters. Similar attempts have been done in
Ref.~\cite{Chacko:1999am,Ibe:2004gh} in anomaly mediation models.
The gaugino mediation~\cite{Kaplan:1999ac} is a variance of the gravity
mediation and known to be a successful framework for solving the flavor
problem. However, since the model relies on the $SW^\alpha W_\alpha$
term for the gaugino masses, the moduli problem and the CP problem
remain unsolved.
In Ref.~\cite{Pomarol:1999ie}, a mixture of anomaly and gauge mediation
is proposed as a solution to the tachyonic slepton problem (see also
\cite{Chacko:2001jt}). The idea is to modify the structure of the
anomaly mediation by introducing an additional light degree of freedom,
$X$. It is claimed that the $\mu$-problem and the tachyonic slepton
problem can be solved by assuming appropriate couplings of $X$ to the
messenger and Higgs fields. However, it is unclear whether such a light
degree of freedom is consistent with cosmological history.

We here propose a simple set-up to address all of the problems. The
assumptions we make are the following: (1)~There is an approximate PQ
symmetry whose small explicit breaking triggers the supersymmetry
breaking. (2)~The Higgs fields and the supersymmetry breaking sector are
directly coupled above an energy scale $\Lambda$. (3)~There are
messenger particles which obtain masses by a VEV of $S$. The effective
Lagrangian written in terms of the Goldstino multiplet $S$ and the MSSM
matter/gauge fields are then given by~\cite{Ibe:2007km}:
\begin{eqnarray}
 K &=& S^\dagger S - \frac{c_S (S^\dagger
  S)^2}{\Lambda^2} \nonumber 
 + \left( \frac{ c_\mu S^\dagger H_u H_d}{\Lambda} + {\rm h.c.} \right)
- 
\frac{ c_H S^\dagger S ( H_u^\dagger H_u + H_d^\dagger H_d ) }{ \Lambda^2
} \nonumber \\
&&
+ \left(
1 - \frac{4 g^4 N_{\rm mess}}{(4 \pi)^4} C_2(R) ( \log |S| )^2 
\right)
\Phi^\dagger \Phi\ 
\ , \nonumber \label{eq:set-up}\\
&&\\
W &=& W_{\rm Yukawa} (\Phi) + m^2 S + w_0 \ , \nonumber \\
&& \nonumber \\
f &=& 
{1 \over 2}
\left(
\frac{1}{g^2}
- \frac{2 N_{\rm mess}}{(4 \pi)^2} \log S
\right) W^\alpha W_\alpha
\nonumber \ .
\end{eqnarray}
The chiral superfield $\Phi$ represents the matter and the Higgs
superfields in the MSSM, and $W_{\rm Yukawa}$ is the Yukawa interaction
terms among them.
We defined $O(1)$ valued coefficients $c_S$, $c_\mu$, and $c_H$. We
normalize the $\Lambda$ parameter so that $c_S = 1$ in the following
discussion.
The parameters $c_H$ and $\Lambda$ take real values whereas $c_\mu$ is a
complex parameter.
We consider the supergravity Lagrangian defined by the above K\"ahler
potential $K$, superpotential $W$, and gauge kinetic function $f$.
This is a closed well-defined system. The linear term of $S$ in the
superpotential represents the source term for the $F$-component of
$S$. 
The last term in the superpotential, $w_0$, is a constant, $|w_0| \simeq m^2
M_{\rm Pl}/\sqrt 3 $, which is needed to cancel the cosmological
constant.
The scalar potential has a minimum at 
\begin{eqnarray}
 \langle S \rangle = {\sqrt 3 \over 6} {\Lambda^2 \over  M_{\rm Pl}}
\label{eq:s-shift}
\end{eqnarray}
which avoids a singularity at $S = 0$ where messenger particles become
massless~\cite{Kitano:2006wz}.

\begin{figure}[t]
\begin{center}
  \includegraphics[height=8.5cm]{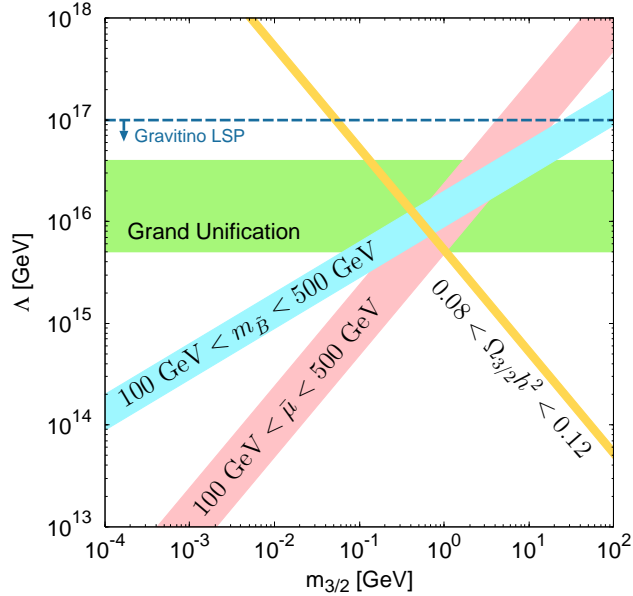}
\end{center}
\caption{Phenomenologically required values of the Higgsino mass $\bar
 \mu$ (with an $O(1)$ ambiguity, see text), the Bino mass $m_{\tilde B}$
 and the gravitino energy density $\Omega_{3/2} h^2$. These three
 quantities have different dependencies on parameters $m_{3/2}$ and
 $\Lambda$. The three bands meet around $m_{3/2} \sim 1$~GeV and
 $\Lambda \sim M_{\rm GUT}$. The quantity $\Omega_{3/2} h^2$ is defined
 in Eq.~(\ref{eq:omega}). It represents the energy density of the
 non-thermally produced gravitinos through the decays of $S$ if $S \to
 hh$ is the dominant decay channel.}
\label{fig:unif}
\end{figure}

We can see the non-trivial success of this framework in
Fig.~\ref{fig:unif}, where we see how $O(1)$~GeV gravitino mass is
selected. The bands of $100~{\rm GeV} < \bar \mu < 500~{\rm GeV}$,
$100~{\rm GeV} < m_{\tilde B} < 500~{\rm GeV}$, and $0.08 < \Omega_{3/2}
h^2 < 0.12$ are shown, where we defined $\bar \mu \equiv m_{3/2} M_{\rm
Pl} / \Lambda$ (typical values of the $\mu$-term) and $\Omega_{3/2} h^2$
by
\begin{eqnarray}
 \Omega_{3/2} h^2 = 0.1
\times 
\left(\frac{m_{3/2}}{500~{\rm MeV}}\right)^{3/2}
\left(\frac{\Lambda}{1 \times 10^{16}~{\rm GeV}}\right)^{3/2}\ .
\label{eq:omega}
\end{eqnarray}
This is a contribution to the matter energy density from non-thermally
produced gravitinos via decays of $S$-condensation in the early
universe~\cite{Ibe:2006rc}.
The Bino mass $m_{\tilde B}$ is the mass of the U(1)$_Y$ gaugino.
Surprisingly, these {\it three} bands meet at $m_{3/2} \sim 1$~GeV and
$\Lambda \sim M_{\rm GUT} \sim 10^{16}$~GeV. The factor of 100 in
Eq.~(\ref{eq:100}) is explained by the ratio $M_{\rm Pl} / M_{\rm GUT}
\sim 100$.

The fact that $\Lambda$ coincides with the unification scale, $M_{\rm
GUT}$, is also quite interesting. It is reasonable that the Higgs fields
are directly coupled to the GUT breaking sector above the GUT scale in
order to achieve the doublet-triplet splitting. It is then also
reasonable to have interaction terms suppressed by the GUT scale after
integrating out heavy fields in the GUT breaking sector.
The same ``cut-off'' scale $\Lambda$ for $S$ and Higgs fields suggests
that the dynamics of GUT breaking is responsible for the supersymmetry
breaking as well. The picture of unification of the Higgs sector, the
supersymmetry breaking sector and the GUT breaking sector naturally
comes out.
Although it sounds like a very ambitious attempt to build a realistic
model to realize this situation, it is quite possible and even very
simple to build such a dream model. For an explicit example of such a
GUT model, see Ref.~\cite{Kitano:2006wm}.

The PQ symmetry plays an essential role in many aspects. It explains (1)
smallness of the $\mu$-term (2) smallness of the supersymmetry breaking
scale (3) absence of the proton-decay operators (4) stability of the
dark matter (5) absence of the CP phase (6) smallness of $\langle S
\rangle$. Especially, (6) smallness of $\langle S \rangle$ by $\langle S
\rangle \propto 1/M_{\rm Pl}$ is important for solving the $\mu$-problem
and also for gravitino cosmology.
This simple framework is free from the $\mu$-problem, flavor problem, CP
problem, proton decay problem or cosmological moduli/gravitino problem.

\section{Low energy phenomenology}

\subsection{Spectrum}
\label{sec:low-energy}

\setcounter{footnote}{0}

The set-up in Eq.~(\ref{eq:set-up}) provides a characteristic spectrum
of the supersymmetric particles. It is different from conventional gauge
or gravity mediation models. Since the Higgs sector directly couples to
the supersymmetry breaking sector at the GUT scale, the soft mass terms
for the Higgs fields are generated at the GUT scale. 
The gaugino masses and sfermion masses are, on the other hand, generated
at the messenger scale. This hybrid feature provides interesting
predictions on the low energy spectrum.

Unfixed parameters in this model are
\begin{eqnarray}
 m_H^2 \left( \equiv {c_H |m^2|^2 \over \Lambda^2} \right)\ , \ \ \ 
\mu \left( \equiv {c_\mu m^2 \over \Lambda} \right)\ ,\ \ \ 
\bar M \left( \equiv {1 \over (4 \pi)^2} {m^2 \over \langle S \rangle}
       \right),\ \ \ 
M_{\rm mess} \left( \equiv k \langle S \rangle \right)\ ,\ \ \ 
N_{\rm mess}
\end{eqnarray}
We take the scale $\Lambda$ to be the unification scale $M_{\rm
GUT}$.

All the soft supersymmetry breaking parameters at the electroweak scale
can be expressed in terms of these five parameters. One combination of
the parameters should be fixed by the condition for the electroweak
symmetry breaking, i.e., $M_Z = 91.2$~GeV. We take the $m_H^2$ parameter
as an output of the calculation. The model parameters are now defined by
$(\mu, M_{\rm mess}, \bar M, N_{\rm mess})$.

\begin{figure}[t]
\begin{center}
  \includegraphics[width=7.5cm]{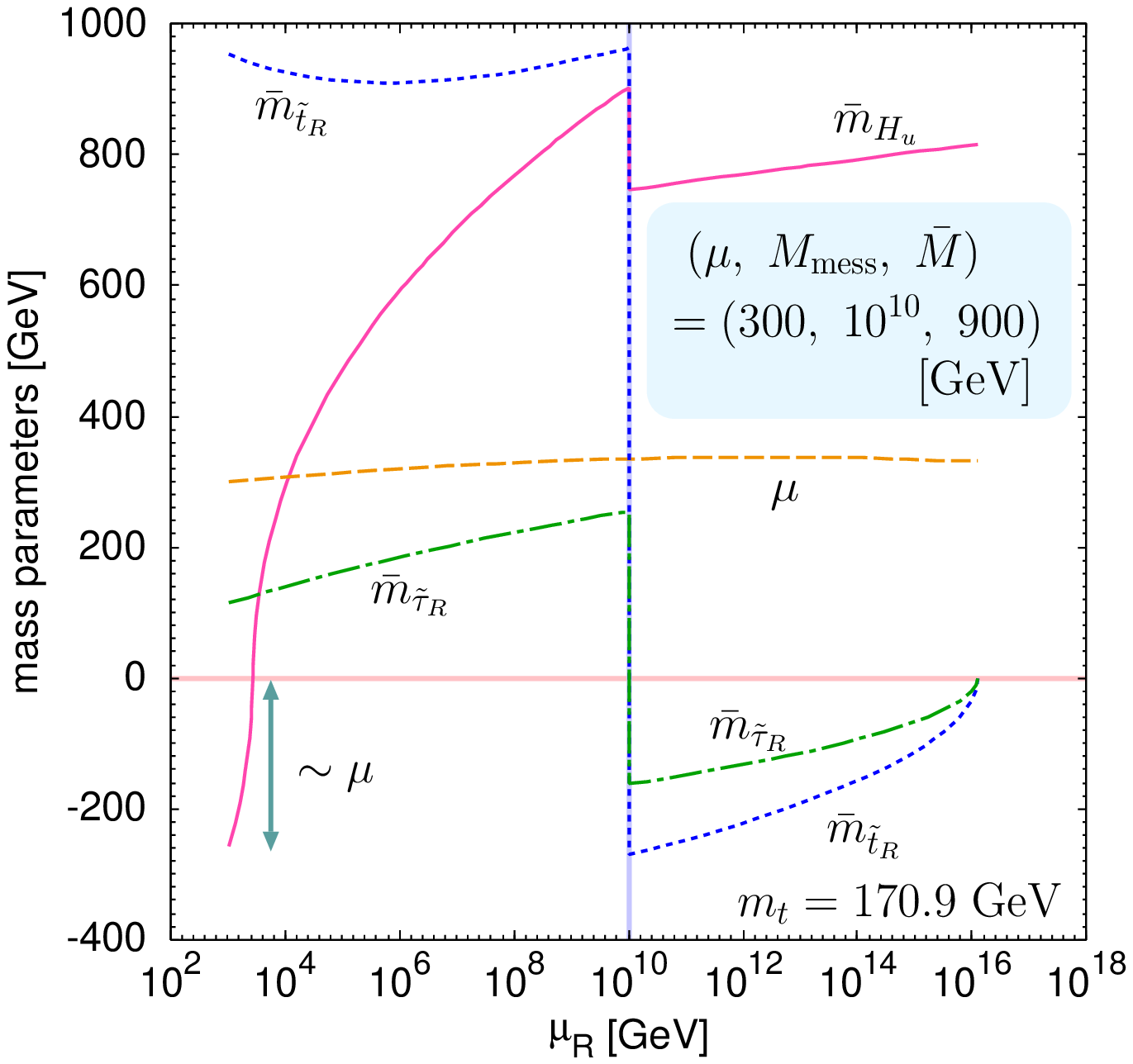}
  \includegraphics[width=7.5cm]{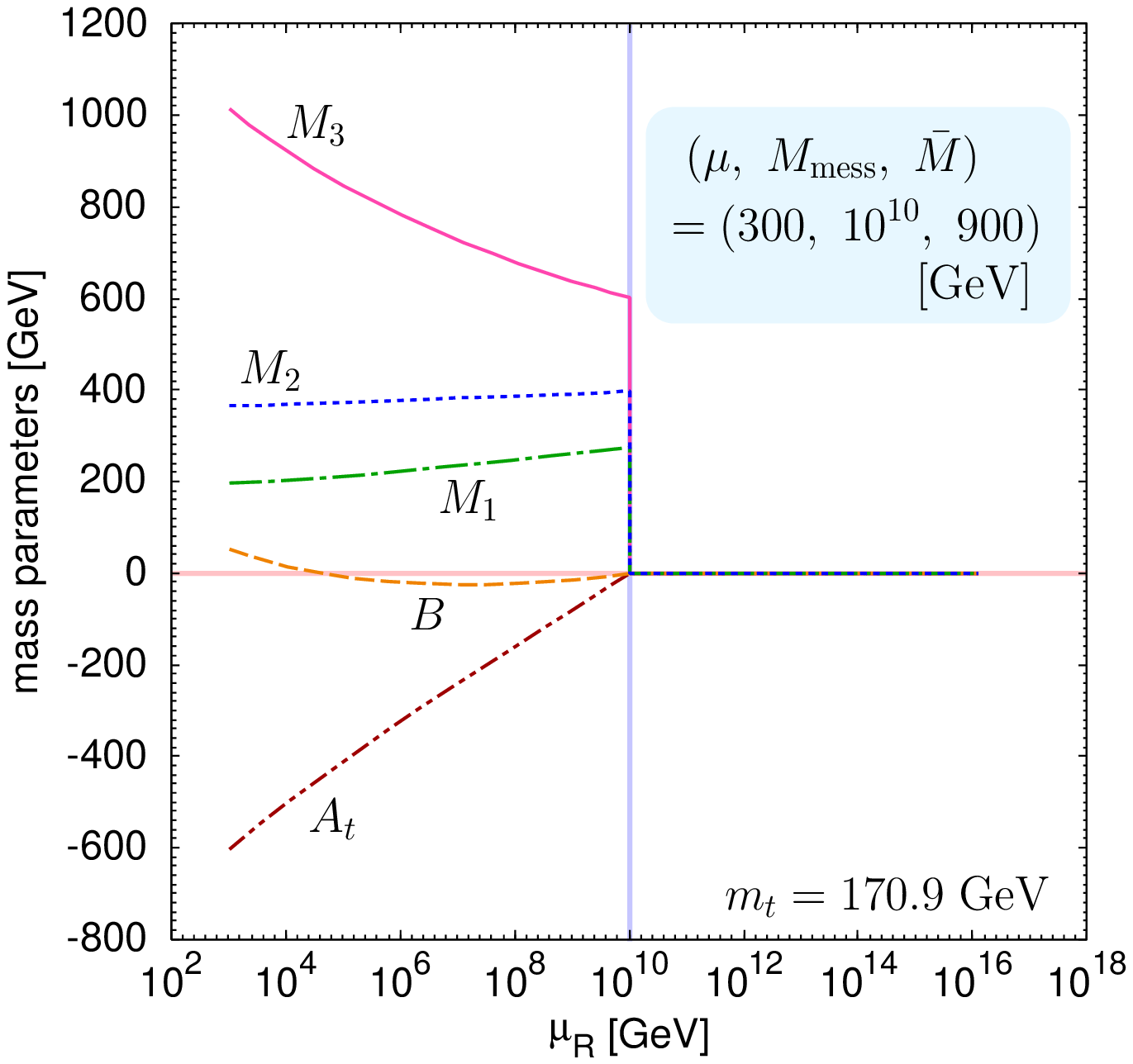}
\end{center}
\caption{The RG evolution of the supersymmetry breaking parameters. We
 take $N_{\rm mess} = 1$.}
 \label{fig:rge}
\end{figure}

We show in Fig.~\ref{fig:rge} an example of the RG evolution of soft
supersymmetry breaking parameters for $(\mu, M_{\rm mess}, \bar M, N_{\rm mess}) =
(300~{\rm GeV}, 10^{10}~{\rm GeV}, 900~{\rm GeV}, 1)$. The horizontal axis $\mu_R$
is the RG scale.
In the left panel of Fig.~\ref{fig:rge}, scalar masses and the
$\mu$-parameter are plotted. We have defined mass parameters $\bar m_{X}
\equiv {\rm sgn} (m_X^2) | m_X^2 |^{1/2}$ for each scalar mass parameter
$m_X^2$. The gaugino masses are generated at the threshold of the
messenger particles.

\subsection{LHC signatures}
\label{sec:lhc}

\setcounter{footnote}{0}

An interesting possibility in this scenario is that the scalar tau
lepton can be the next to lightest supersymmetric particle (NLSP) due to
the negative contribution from the one-loop running above the messenger
scale. This situation is actually well-motivated from the discussion of
the ultra-violet completion of the framework~\cite{Ibe:2007km}.
If the stau is the NLSP, the lifetime is of $O(1000)$~seconds with our
assumption of the $O(1)$~GeV gravitinos.
The LHC signals with such a long-lived stau will be quite different from
ones with the usual assumption of the neutralino LSP.

We study the LHC signals with the parameter set we have used in
Fig.~\ref{fig:rge} where the stau is the NLSP.  The stau will be
produced at the LHC mainly from the neutralino decays $\chi_i^0 \to
\tilde \tau \tau$, and most of them reach to the muon system. Therefore,
we can reconstruct the four-momenta of staus event by event basis with a
good accuracy~\cite{Ambrosanio:2000ik}. The reconstruction of neutralino
masses is also possible by looking at the edge of the invariant mass of
the stau and a tau-jet, $M_{\tilde \tau \tau}$~\cite{Hinchliffe:1998ys}
(see \cite{Ellis:2006vu} for a recent study in minimal supergravity
model).

\begin{figure}[t]
\begin{center}
  \includegraphics[height=7.4cm]{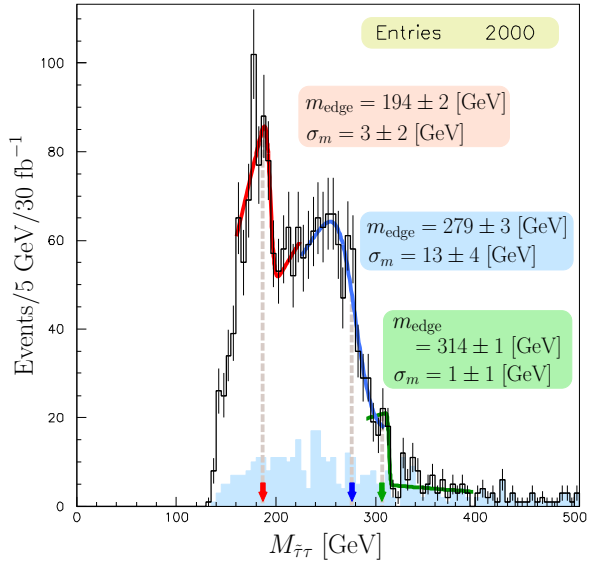}
\end{center}
\caption{The distribution of the invariant mass $M_{\tilde \tau \tau}$.
The shaded histogram shows the events with a mis-identified $\tau$-jet
which is simulated by assuming a mis-tagging probability of a
non-$\tau$-labelled jet to be 1\%.
The small allows and dashed lines denote the input values of three
neutralino masses.  Three curves are fitting functions of three
endpoints which correspond to the endpoints of $\chi_{1,2,3}^{0}$ from
left to right, respectively. The third endpoint is statistically not
very significant. We have used HERWIG 6.50~\cite{Corcella:2002jc} for
event generation, TAUOLA 2.7~\cite{Jadach:1993hs} for simulation of tau
decays and AcerDET 1.0~\cite{Richter-Was:2002ch} for the detector
simulation.}
\label{fig:mstautau}
\end{figure}

Fig.~\ref{fig:mstautau} shows the distribution of the invariant
mass.
We can clearly see the edges at the input neutralino masses. We can
confirm/exclude the model by testing if relations among stau and
neutralino masses are consistent with the model predictions.

\section*{Acknowledgements}

RK thanks the organizers of the Summer Institute 2007 for the invitation
and the nice meeting in Fuji-Yoshida. He especially thanks Prof.~Jisuke
Kubo and Prof.~Haruhiko Terao for their hospitality during his (and his
family's) stay in Kanazawa and Fuji-Yoshida.

\noindent
The Summer Institute 2007 is sponsored by 
JSPS Grant-in Aid for Scientific Research (B)
No. 16340071 and also partly by the Asia 
Pacific Center for Theoretical Physics, APCTP.


\end{document}